\documentclass[compactaffiliation]{Interspeech}

\interspeechcameraready

\usepackage{svg}
\usepackage{lipsum}
\usepackage{orcidlink}
\usepackage{colortbl}
\usepackage{booktabs}
\usepackage{multicol}
\usepackage{arydshln}
\usepackage{multirow}
\usepackage{adjustbox}
\usepackage{threeparttable}

\title{On the Relevance of Clinical Assessment Tasks for the Automatic Detection of Parkinson’s Disease Medication State from Speech}

\author[affiliation={1,2}]{David}{Gimeno-Gómez}
\author[affiliation={2}]{Rubén}{Solera-Ureña}
\author[affiliation={2}]{Anna}{Pompili}
\author[affiliation={1}]{Carlos-D.}{Martínez-Hinarejos}
\author[affiliation={4,5}]{Rita}{Cardoso}
\author[affiliation={4,6}]{Isabel}{Guimarães}
\author[affiliation={4,5}]{Joaquim J.}{Ferreira}
\author[affiliation={2,3}]{Alberto}{Abad}

\affiliation{PRHLT}{Universitat Politècnica de València}{Spain} 
\affiliation{INESC-ID}{Lisbon}{Portugal}
\affiliation{Instituto Superior Técnico}{Universidade de Lisboa}{Portugal}
\affiliation{Laboratory of Clinical Pharmacology and Therapeutics}{Faculdade de Medicina, Universidade de Lisboa}{Portugal}
\affiliation{CNS - Campus Neurológico}{Torres Vedras}{Portugal}
\affiliation{Alcoitão Health School of Sciences}{Santa Casa da Misericórdia de Lisboa}{Portugal}

\email{dagigo1@dsic.upv.es}


\usepackage{comment}

\begin{document}

\maketitle

\begin{abstract}
The automatic identification of medication states of Parkinson's disease (PD) patients can assist clinicians in monitoring and scheduling personalized treatments, as well as studying the effects of medication in alleviating the motor symptoms that characterize the disease. This paper explores speech as a non-invasive and accessible biomarker for identifying PD medication states, introducing a novel approach that addresses this task from a speaker-independent perspective. While traditional machine learning models achieve competitive results, self-supervised speech representations prove essential for optimal performance, significantly surpassing knowledge-based acoustic descriptors. Experiments across diverse speech assessment tasks highlight the relevance of prosody and continuous speech in distinguishing medication states, reaching an F1-score of 88.2\%. These findings may streamline clinicians' work and reduce patient effort in voice recordings.
\end{abstract}

\section{Introduction}

Parkinson’s disease (PD) is a progressive neurodegenerative disorder that primarily affects movement control and speech production \cite{jankovic2008parkinson}. Patients are faced with articulatory deficits, leading to dysarthria, and voice disorders, frequently perceived as dysprosody (resulting in voice modulation and intensity impairments) \cite{skodda2011intonation}. Within the course of the disease, speech decline and communication disorders worsen with a nonlinear progression \cite{hilker2005nonlinear}. In this context, monitoring of PD patients becomes essential for a more targeted medical follow-up and treatment. In particular, the identification of the periods in which motor symptoms are mitigated by the effect of medication intake (ON state), and the periods with motor complications (OFF state), may allow clinicians to schedule and adjust medication intake, monitoring and analyzing the course of the disease over time, and investigating their correlation with motor fluctuations.

Clinical literature has shown controversial results on the effects of Levodopa on speech, both for early and late-stage PD subjects. While some studies reported favorable results \cite{nakano1973speech,wolfe1975speech}, others observed no improvements or even detrimental effects of pharmacological treatment on speech \cite{skodda2011intonation,plowman2009perceptual}. Regarding the computational analysis of pathological speech, a primary research focus has been the automatic diagnosis of PD condition from speech. Early works employed traditional methods, such as knowledge-based acoustic descriptors and conventional machine learning techniques \cite{orozco2016automatic,vergara2017aging,lahoti2022long,orozco2015voice,pompili2017automatic}, while more recent studies have explored deep learning-based approaches \cite{wagner2023pathological,laquatra2024exploiting,escobar2024foundation,gallo2024levels}. Over the past few years, self-supervised (SSL) speech representations have gained attention for their effectiveness in pathological speech tasks, including PD detection \cite{wagner2023pathological,laquatra2024exploiting,favaro2023interpretable}. However, there is a significative lack in the literature concerning systematic research on the computational identification or monitoring of PD patient's medication states through speech. A noteworthy exception is the work of Pompili et al. \cite{pompili2020medication}, where a speaker-dependent approach that combined eGeMAPS features with feed-forward networks is proposed. This speaker-dependent approach, however, limits its applicability in real-world scenarios, as it implies the development of a new, personalized model for each new individual subject.

All these factors motivate our present work, which presents the first approach for the automatic, speaker-independent identification of medication states of PD patients from speech. We explore the combination of two types of speech representations --- knowledge-based (eGeMAPS \cite{eyben2016egemaps}) features and SSL (Wav2Vec2.0 \cite{babu2202wav2vec}) speech embeddings ---, as well as two model architectures: support vector machines (SVMs) and attention-based deep neural networks (A-DNNs). Despite the inherent complexity of the task, the experimental assessment on the European Portuguese FraLusoPark corpus \cite{pinto2016fralusopark} shows robust performance, achieving an F\textsubscript{1}-score of 88.2\%. Compared to \cite{pompili2020medication}, this work presents the following key contributions:

\begin{itemize}

    \item  We devise here a speaker-independent approach that offers two main practical advantages: i) it benefits from richer, cross-speaker information in larger datasets to develop more robust systems, and ii) in deployment, it eliminates the need to develop personalized models for each new patient.

    \item We show the potential of SSL speech representations in capturing the nuances of distinct medication states, significantly outperforming traditional knowledge-based features, even without the need for fine-tuning.
    
    \item We show that traditional machine-learning classifiers, such as SVMs, achieve competitive results compared to more demanding attention-based deep architectures, which reduces computational requirements and makes our approach more suitable for real-world screening scenarios.

    \item We perform a comprehensive experimental assessment for several speech tasks commonly used in clinical practice. In line with the literature \cite{frota2021dyspros}, our analyses highlight the relevance of prosodic cues in continuous speech tasks for the identification of medication states in PD patients.
    
\end{itemize}

\vspace{-0.1cm}
\section{Data Material}

The European Portuguese FraLusoPark corpus \cite{pinto2016fralusopark} is composed of 158 speakers, 74 healthy controls and 84 PD patients. A unique feature of this corpus is the inclusion of a study specifically designed to analyze the effects of medication in PD patients. To this end, participants diagnosed with PD were recorded twice: at least 12 hours after withdrawal of all anti-Parkinsonian drugs (OFF), and at least 1 hour after the administration of the usual medication (ON). In each medication state, participants were evaluated by expert neurologists and assessed by speech-language therapists (SLTs) according to the Movement Disorder Society-Unified Parkinson’s Disease Rating Scale (MDS-UPDRS-III) \cite{goetz2008updrs}. The type and dosage of pharmacological treatment vary for each individual participant.

For this study, patients with atypical Parkinsonian disorders or suffering from cognitive decline or severe depression were excluded. Thus, the subset considered in this work consists of 74 PD patients, 36 males and 38 females. Their average ages are 66.9{\scriptsize$\pm$8.6} and 64.6{\scriptsize$\pm$11.6} years, respectively, while their average MDS-UPDRS-III scores are 47.3{\scriptsize$\pm$18.2} and 41.5{\scriptsize$\pm$14.8}. Notably, after medication intake, 25 out of 74 participants experienced changes in their MDS-UPDRS-III assessments, with two of them worsening their scores. This outcome, along with the variation in treatments among subjects, highlights the diversity in individual responses to medication and its effects on PD symptoms --- further underscoring the complexity of this task.

The protocol included nine speech assessment tasks designed to be administered by SLTs in a fixed order of increasing complexity. Each participant performed one recording sample for each task, resulting in a total of 666 samples for each condition (ON/OFF). Table~\ref{tab:tasks} provides a brief description of the tasks.


\begin{table}[!htbp]
\centering
\scriptsize

\caption{Description of the speech tasks in FraLusoPark.}
\label{tab:tasks}

\begin{adjustbox}{max width=0.95\columnwidth}

\begin{threeparttable}
\begin{tabular}{ll}
 \toprule
 
  \textbf{Speech Task} & \textbf{Description} \\ \midrule \midrule
 \textbf{A} & {\parbox{7cm} {steady vowel /a/ phonation (at a comfortable pitch and loudness), repeated three times}}  \\ \midrule 
 \textbf{MPT} & {\parbox{7cm} {maximum phonation time (vowel /a/ sustained as long as possible on one deep breath at a comfortable pitch and loudness), repeated twice}} \\ \midrule
 \textbf{DDK} & {\parbox{7cm} {oral diadochokinesia (repetition of the pseudoword \textit{pataka}) at a fast rate for 30s }}\\ \midrule
 \textbf{WORDS} & {\parbox{7cm} { reading aloud of 10 words created by adapting the intelligibility part of V.2 of the Frenchay Dysarthria Assessment (FDA-2) \cite{cardoso2017frenchay} }} \\ \midrule
 \textbf{SENT} & {\parbox{7cm} { reading aloud of 10 sentences created by adapting the intelligibility part of V.2 of the FDA-2 
 \cite{cardoso2017frenchay} }} \\ \midrule
 \textbf{PROS-SENT} & {\parbox{7cm} { reading aloud a set of 20 sentences with specific language-dependent prosodic properties }} \\ \midrule
 \textbf{TEXT} & {\parbox{7cm} { reading aloud of a short text (\textit{‘The North Wind and the Sun’}, European Portuguese adaptation) }} \\ \midrule
 \textbf{FROG} & {\parbox{7cm} { storytelling speech guided by visual stimuli (pictures from the wordless story \textit{‘Frog, Where are you?'}) }} \\ \midrule
 \textbf{CONVERS} & {\parbox{7cm} { free conversation for around 3 minutes }} \\ 
 \bottomrule

\end{tabular}
\end{threeparttable}
\end{adjustbox}
\vspace{-0.4cm}
\end{table}

\vspace{-0.1cm}
\section{Methodology}

This work is inspired by two well-established frameworks in the field of automatic analysis of pathological speech: i) the traditional approach, which relies on knowledge-based acoustic descriptors and conventional machine learning techniques, and ii) a more recent paradigm that leverage SSL speech representations and attention-based deep learning architectures. In particular, our approach explores the combination of two types of speech representations --- knowledge-based (eGeMAPS) feature descriptors and SSL (Wav2Vec2.0) speech embeddings --- as well as two model architectures: SVMs and A-DNNs.

\vspace{-0.2cm}
\subsection{Feature Extraction}

Before feature extraction, speech samples were first downsampled to 16 kHz. Then, we adopted the automatic approach described in \cite{pompili2020medication} to remove speech segments corresponding to silent pauses and interventions (instructions) of the SLTs. Clean speech samples underwent EBU R128\footnote{\url{https://tech.ebu.ch/publications/r128/}} loudness normalization.

\noindent\textbf{eGeMAPS} \cite{eyben2016egemaps} are knowledge-based speech features widely recognized for their utility in various paralinguistic tasks \cite{zhao2019exploring,mallol2024multi}, including pathological speech \cite{orozco2015voice}. eGeMAPS were extracted at two different levels using openSMILE \cite{eyben2010opensmile}: 25 low-level descriptors (LLDs) at a frame rate of 10ms, and 88 functional statistics that aggregate these LLDs at the utterance-level.

\noindent\textbf{Wav2Vec2.0} \cite{babu2202wav2vec} self-supervised speech representations have shown to be effective in various pathological speech analysis tasks \cite{wagner2023pathological,laquatra2024exploiting}. Here, we used the 300M-parameter, 24-layer, multilingual (128 languages, including Portuguese) version\footnote{\url{https://pytorch.org/audio/main/generated/torchaudio.pipelines.WAV2VEC2_XLSR_300M.html}}. Inspired by prior works in this field \cite{favaro2023interpretable,purohit2025layeradapt} and existing literature on layer-wise analysis of SSL-based foundational models \cite{pasad2021layerwise}, we extracted 1024-dimensional embeddings from the 7th intermediate encoder layer. Features were also computed at two levels: the original temporal embedding sequence sampled every 20ms, and an utterance-level representation derived by computing the mean, standard deviation, kurtosis, and skewness across time, resulting in a 4$\times$1024-dimensional matrix.

Finally, speech features were standardized similarly to Kovac et al. \cite{kovac2024exploring} using the median and standard deviation computed over the ON-state training samples. This process considers the ON-state as the reference population, as dopaminergic medication (although subject to individual variability) is the standard treatment to alleviate patients' symptoms~\cite{murakami2023PDdrugtherapy}\footnote{Notice that no significant differences were observed using the entire dataset to compute those statistics or without any normalization.}.

\vspace{-0.2cm}
\subsection{Model Architectures and Training}

\noindent\textbf{Linear Support Vector Machines (SVM)} were implemented using Scikit-Learn\footnote{\url{https://scikit-learn.org/dev/modules/generated/sklearn.svm.LinearSVC.html}}. Models were trained using the utterance-level speech features. Principal Component Analysis (PCA) was applied for dimensionality reduction.

\noindent\textbf{Attention-based Deep Neural Networks (A-DNN)}. This architecture comprises three main components: (i) a linear layer that projects input features into a 1024-dimensional latent space, (ii) a self-attention mechanism to capture long-term temporal relationships, and (iii) a sample-level classification module consisting of layer normalization, average pooling, a non-linear Swish activation function, 
and a final linear layer. Models were trained using the frame-level speech features.

\noindent\textbf{5-fold Nested Cross-Validation} was employed for reliable training and evaluation. To prevent overfitting, an outer loop is used to compute an estimate of the model's performance, while the inner loop is used to perform hyperparameter optimization. Dataset splitting was designed to ensure speaker-independence and balanced gender representation. To address variability often associated with small datasets, each experiment was repeated 5 times with different random seeds. The final performance is reported as the average F\textsubscript{1}-score across folds and runs.

\noindent\textbf{Training Settings}. Experiments with the A-DNN model were conducted on GeForce GTX TITAN X GPUs with 12GB memory. Based on a preliminary hyperparameter search, we opted for a fixed setup, with the AdamW optimizer, a learning rate of 0.0004, cosine-scheduled decay over 5 epochs, and a batch size of 8 samples. For the SVM-based experiments, hyperparameters (number of principal components for PCA $\in$ [16, 32, 64], and SVM regularization parameter $C$ $\in$ [0.01, 0.1, 1, 10]) were selected via cross-validation as described above.

\vspace{-0.2cm}
\subsection{Data Grouping Strategies for Training}

To investigate the performance and robustness of our methods from multiple perspectives, considering not only the amount of training data but also the degree of task specialization, three distinct data grouping strategies were explored in our experiments.

\noindent\textbf{Task-Specific}. This strategy involves training and evaluating a separate model for each speech assessment task, using exclusively the data corresponding to its specific target task. This represents the most constrained scenario in terms of generalizability, enabling us to assess the effectiveness and robustness of our approach under very scarce data settings.

\noindent\textbf{Task-Grouping.} In this strategy, different speech tasks are grouped together based on their similarities. Specifically, we defined the following groups: (i) \texttt{A}+\texttt{MPT}, (ii) \texttt{WORDS}+\texttt{DDK}, (iii) \texttt{SENT}+\texttt{PROS-SENT}+\texttt{TEXT}, and (iv) \texttt{FROG}+\texttt{CONVERS}. A single model is trained and evaluated with all the data from the tasks in a given group. This scenario is a compromise between task specialization and generalization, allowing our approach to leverage shared patterns within related tasks to potentially enhance performance, particularly for less-resourced speech tasks.

\noindent\textbf{Task-Independent}. In this approach, a single model is trained and evaluated using the data from all the assessment tasks. This represents our most generalizable scenario, enabling us to study how effectively our approach can benefit from larger amounts of training data and adapt to diverse speech assessment tasks.

\vspace{-0.1cm}
\section{Results \& Discussion}

A comprehensive overview of our experimental results for automatic, speaker-independent PD medication state identification from speech is presented in Table \ref{tab:results}, showing the outcomes for the three data grouping strategies and different combinations of model architectures and speech features. Results are presented at the utterance level and were computed individually for each speech assessment task, using in each case the corresponding Task-\{Specific/Grouping/Independent\} model. Further analysis across gender (Table \ref{tab:gender}) and dysarthria severity levels are also presented and discussed.

\begin{table*}[!ht]
\centering
\small
\caption{Results (F\textsubscript{1}-score - \%) for speaker-independent medication state identification from speech in the FraLusoPark corpus. Best results for each model and speech task highlighted in \textbf{bold}, best overall results for each speech task in \colorbox{gray!20}{gray shading}.}
\label{tab:results}
\vspace{-0.2cm}
\begin{adjustbox}{max width=1.0\textwidth}
\begin{threeparttable}
\begin{tabular}{lccccccccccccccccccccc}
 \toprule
 
 \multirow{2}{*}[-3pt]{\textbf{Training Strategy}} & & \multirow{2}{*}[-3pt]{\textbf{Features}} & & \multicolumn{17}{c}{\textbf{Speech Assessment Tasks}} \\ \cmidrule{5-21}
 & & & & \textbf{A} & & \textbf{MPT} & & \textbf{WORDS} & & \textbf{DDK} & & \textbf{SENT} & & \textbf{PROS-SENT} & & \textbf{TEXT} & & \textbf{FROG} & & \textbf{CONVERS}  \\ \midrule
 

 \multicolumn{22}{c}{\textit{Support Vector Machine (SVM)}} \\ \midrule\midrule

 \multirow{2}{*}[0pt]{\textbf{Task-Specific}} & &     \textit{eGeMAPS} & & 46.9{\tiny $\pm$0.3} & & 54.7{\tiny $\pm$0.0} & & 54.1{\tiny $\pm$0.0} & & 48.6{\tiny $\pm$0.0} & & 49.2{\tiny $\pm$0.0} & & 62.8{\tiny $\pm$0.0} & & 62.8{\tiny $\pm$0.0} & & 58.7{\tiny $\pm$0.3} & & 50.4{\tiny $\pm$0.8} \\ 
 & & \textit{Wav2Vec2.0} & & 53.2{\tiny $\pm$2.5} & & 54.0{\tiny $\pm$1.0} & & 55.5{\tiny $\pm$1.4} & & 55.8{\tiny $\pm$1.7} & & 55.6{\tiny $\pm$2.2} & & \cellcolor[gray]{0.9}\textbf{88.2{\tiny $\pm$1.5}} & & 68.5{\tiny $\pm$2.0} & & 78.9{\tiny $\pm$2.4} & & 59.0{\tiny $\pm$1.2} \\

 \noalign{\vskip 0.5ex} \hdashline \noalign{\vskip 0.5ex}

 \multirow{2}{*}[0pt]{\textbf{Task-Grouping}} & & \textit{eGeMAPS} & & 50.3{\tiny $\pm$0.3} & & \cellcolor[gray]{0.9}\textbf{56.1{\tiny $\pm$1.4}} & & 42.8{\tiny $\pm$0.8} & & 44.0{\tiny $\pm$2.2} & & 48.3{\tiny $\pm$1.6} & & 56.8{\tiny $\pm$0.4} & & 53.0{\tiny $\pm$2.0} & & 61.6{\tiny $\pm$0.6} & & 55.8{\tiny $\pm$0.5} \\ 
 & & \textit{Wav2Vec2.0} & & \textbf{53.9{\tiny $\pm$1.6}} & & 54.0{\tiny $\pm$1.5} & & \textbf{55.7{\tiny $\pm$2.4}} & & \textbf{58.7{\tiny $\pm$1.7}} & & \textbf{61.1{\tiny $\pm$2.1}} & & 76.5{\tiny $\pm$1.7} & & \cellcolor[gray]{0.9}\textbf{71.5{\tiny $\pm$2.1}} & & \cellcolor[gray]{0.9}\textbf{80.1{\tiny $\pm$1.1}} & & \textbf{60.7{\tiny $\pm$0.8}}\\

 \noalign{\vskip 0.5ex} \hdashline \noalign{\vskip 0.5ex}
 
 \multirow{2}{*}[0pt]{\textbf{Task-Independent}} & & \textit{eGeMAPS} & & 44.4{\tiny $\pm$1.6} & & 51.7{\tiny $\pm$0.5} & & 48.8{\tiny $\pm$0.2} & & 45.5{\tiny $\pm$0.6} & & 48.8{\tiny $\pm$1.6} & & 54.4{\tiny $\pm$1.0} & & 57.2{\tiny $\pm$1.6} & & 50.7{\tiny $\pm$0.3} & & 50.5{\tiny $\pm$1.0} \\ 
 & & \textit{Wav2Vec2.0} & & 53.2{\tiny $\pm$1.1} & & 53.1{\tiny $\pm$1.4} & & 55.7{\tiny $\pm$2.4} & & 58.4{\tiny $\pm$1.9} & & 54.6{\tiny $\pm$1.4} & & 63.5{\tiny $\pm$1.7} & & 55.3{\tiny $\pm$1.9} & & 65.5{\tiny $\pm$0.8} & & 59.0{\tiny $\pm$1.2} \\ \midrule


 \multicolumn{22}{c}{\textit{Attention-based Deep Neural Network (A-DNN)}} \\ \midrule\midrule

 \multirow{2}{*}[0pt]{\textbf{Task-Specific}} & &     \textit{eGeMAPS} & & 52.5{\tiny $\pm$2.9} & & 52.5{\tiny $\pm$2.7} & & 47.2{\tiny $\pm$4.5} & & 52.3{\tiny $\pm$2.9} & & 51.7{\tiny $\pm$1.2} & & 49.9{\tiny $\pm$1.1} & & 50.4{\tiny $\pm$2.6} & & 53.1{\tiny $\pm$2.2} & & 50.0{\tiny $\pm$1.4} \\ 
 & & \textit{Wav2Vec2.0} & & 52.1{\tiny $\pm$3.1} & & 54.6{\tiny $\pm$2.8} & & 62.4{\tiny $\pm$2.2} & & 58.1{\tiny $\pm$1.1} & & 60.2{\tiny $\pm$0.8} & & \textbf{83.8{\tiny $\pm$0.7}} & & \textbf{70.4{\tiny $\pm$1.2}} & & \textbf{78.8{\tiny $\pm$0.7}} & & 67.1{\tiny $\pm$2.8} \\

 \noalign{\vskip 0.5ex} \hdashline \noalign{\vskip 0.5ex}
 
 \multirow{2}{*}[0pt]{\textbf{Task-Grouping}} & & \textit{eGeMAPS} & & 55.9{\tiny $\pm$1.8} & & 52.7{\tiny $\pm$1.4} & & 45.8{\tiny $\pm$1.0} & & 49.6{\tiny $\pm$2.2} & & 50.0{\tiny $\pm$1.3} & & 49.4{\tiny $\pm$2.4} & & 51.4{\tiny $\pm$1.9} & &  51.7{\tiny $\pm$3.2}  & & 48.7{\tiny $\pm$1.2} \\ 
 & & \textit{Wav2Vec2.0} & & 52.0{\tiny $\pm$2.5} & & 53.9{\tiny $\pm$2.7} & & 56.9{\tiny $\pm$3.0} & & \cellcolor[gray]{0.9}\textbf{61.7{\tiny $\pm$2.7}} & & 63.8{\tiny $\pm$0.9} & & 80.7{\tiny $\pm$1.9} & & 69.2{\tiny $\pm$2.4} & & 73.2{\tiny $\pm$1.1} & & 65.4{\tiny $\pm$0.5} \\ 

 \noalign{\vskip 0.5ex} \hdashline \noalign{\vskip 0.5ex}

 \multirow{2}{*}[0pt]{\textbf{Task-Independent}} & & \textit{eGeMAPS} & & 50.4{\tiny $\pm$1.4} & & 49.1{\tiny $\pm$0.9} & & 48.8{\tiny $\pm$2.3} & & 46.4{\tiny $\pm$5.5} & & 47.5{\tiny $\pm$2.3} & & 49.6{\tiny $\pm$3.7} & & 49.2{\tiny $\pm$2.6} & & 49.5{\tiny $\pm$1.9} & & 48.5{\tiny $\pm$1.9} \\ 
 & & \textit{Wav2Vec2.0} & & \cellcolor[gray]{0.9}\textbf{57.5{\tiny $\pm$2.7}} & & \textbf{54.7{\tiny $\pm$5.2}} & & \cellcolor[gray]{0.9}\textbf{63.3{\tiny $\pm$1.6}} & & 60.2{\tiny $\pm$1.7} & & \cellcolor[gray]{0.9}\textbf{65.1{\tiny $\pm$2.1}} & & 77.6{\tiny $\pm$0.5} & & 67.8{\tiny $\pm$1.4} & & 72.6{\tiny $\pm$0.9} & & \cellcolor[gray]{0.9}\textbf{68.2{\tiny $\pm$1.7}} \\ \bottomrule 

\vspace{-0.9cm}
\end{tabular}
\end{threeparttable}
\end{adjustbox}
\end{table*}

\noindent\textbf{Speech Assessment Tasks.} One of the first conclusions we can infer from the results reported in Table \ref{tab:results} is the inherent complexity of addressing the speaker-independent identification of medication states in PD patients through speech analysis. This is evidenced by the weak performance observed in many speech assessment tasks, with F\textsubscript{1}-scores under or around 60\%.  Interestingly, performance improves significantly as tasks involve more continuous and semi-spontaneous speech production. Indeed, tasks such as \texttt{PROS-SENT}, \texttt{TEXT}, and \texttt{FROG} achieve F\textsubscript{1}-score performances above 70\%, underscoring the value of natural speech for the assessment of PD medication state. Furthermore, our best performing model, using SSL-based Wav2Vec2.0 features and the SVM classifier, achieves an F\textsubscript{1}-score of 88.2\% using the task-specific strategy with the \texttt{PROS-SENT} assessment task. This suggests that prosody may be a specially relevant speech biomarker to discern between medication states in PD.

\noindent\textbf{Speech Features.} We can observe that the knowledge-based eGeMAPS acoustic descriptors are outperformed by SSL-based Wav2Vec2.0 embeddings in almost all scenarios, with particularly significant differences in the top performing ones (\texttt{PROS-SENT}, \texttt{TEXT}, and \texttt{FROG}). It is noteworthy that eGeMAPS, despite including prosodic features, do not show comparable improvements in prosody-related tasks. While effective for PD medication state detection in controlled, speaker-dependent scenarios \cite{pompili2020medication}, we hypothesize their low dimensionality may limit their capability to generalize to more complex settings. There, SSL embeddings trained on large spontaneous speech corpora from many speakers may better capture prosodic variability. This finding supports the potential and growing application of SSL representations in the context of pathological speech analysis. However, despite their promising performance, they often function as opaque, black-box models, unlike traditional features. This lack of interpretability is a critical limitation for clinical applications, where understanding the model’s decisions is crucial. Consequently, there is a growing need for more interpretable SSL-based speech representations \cite{gimeno2025unveiling}.

\noindent\textbf{Model Architecture.} Traditional linear SVMs achieve competitive results compared to A-DNNs, except for the \texttt{WORDS} and \texttt{CONVERS} tasks. The results for the task-independent strategy show that, in general, A-DNNs benefit from more training data, independently of their heterogeneous nature. However, for the \texttt{PROS-SENT}, \texttt{TEXT}, and \texttt{FROG} speech tasks, SVMs are capable to close the gap and even outperform A-DNNs when trained with smaller and more homogeneous datasets. In the \texttt{PROS-SENT} task, SVMs outperform the A-DNN models by a significant margin and achieve the overall best result with an F\textsubscript{1}-score of 88.2\%. For the \texttt{CONVERS} task, we hypothesize that the longer duration of speech samples makes the processing of temporal dependencies more critical, which may explain why A-DNNs are more effective in this case.

\noindent\textbf{Data Grouping Strategy.} In this work, we addressed the dilemma of training data size versus heterogeneity that frequently appears in less-resourced tasks. More data --- potentially corresponding to significantly different speech tasks --- does not necessarily lead to improved performance. This insight motivated the definition of the three distinct data grouping strategies for training. Our experimental results show that, for the less-performant speech tasks \texttt{A}, \texttt{MPT}, \texttt{WORDS}, \texttt{DDK}, and \texttt{SENT}, differences between the three data grouping strategies are minor or even negligible. For \texttt{PROS-SENT}, \texttt{TEXT}, and \texttt{FROG}, both SVMs and A-DNNs architectures show better results in the task-specific or task-grouping settings, suggesting that the use of more homogeneous, continuous, and prosodically rich speech data favors the automatic identification of PD patients' medication state. In particular, the \texttt{PROS-SENT} speech assessment task emerges itself as a very promising, low-cost option for this task using a task-dependent setting.

\begin{table}[!htbp]
\centering
\scriptsize
\caption{Gender-based performance (F\textsubscript{1}-score - \%) for the \texttt{PROS-SENT} task and the task-specific data grouping strategy.}
\label{tab:gender}
\vspace{-0.2cm}

\begin{adjustbox}{max width=0.9\columnwidth}
\begin{threeparttable}
\begin{tabular}{lccccccc}
 \toprule
 
 & \multirow{2}{*}[-2pt]{\textbf{\begin{tabular}[c]{@{}c@{}}Model\\Architecture\end{tabular}}} & & \multicolumn{5}{c}{\textbf{Gender}} \\ \cmidrule{4-8}
 & & & \textit{All} & & \textit{Male} & & \textit{Female} \\ \midrule
 
 \multirow{2}{*}[0pt]{\textbf{\begin{tabular}[c]{@{}c@{}}Gender\\Independent\end{tabular}}} & \textit{SVM} & & 88.2{\tiny $\pm$1.5} & & 88.9{\tiny $\pm$2.0} & & 87.6{\tiny $\pm$1.8} \\
  & \textit{A-DNN} & & 83.8{\tiny $\pm$0.7} & & 83.3{\tiny $\pm$2.2} & & 84.1{\tiny $\pm$2.3} \\ \midrule

 \multirow{2}{*}[0pt]{\textbf{\begin{tabular}[c]{@{}c@{}}Gender\\Dependent\end{tabular}}} & \textit{SVM} & & 86.6{\tiny $\pm$0.7} & & 88.0{\tiny $\pm$1.7} & & 85.3{\tiny $\pm$1.5} \\
  & \textit{A-DNN} & & 78.0{\tiny $\pm$2.4} & & 76.1{\tiny $\pm$3.0} & & 79.9{\tiny $\pm$2.3} \\

 \bottomrule

\end{tabular}
\end{threeparttable}
\end{adjustbox}
\vspace{-0.65cm}
\end{table}

\noindent\textbf{Gender-based Analysis.} Here, we assess the robustness of our approach with respect to the subjects' gender. For conciseness, we limit our analysis to our best-performing configuration that considers SVM and A-DNN classifiers trained and evaluated on the \texttt{PROS-SENT} speech task using SSL-based features. Table~\ref{tab:gender} shows again the performance of the system described above (\texttt{Gender-Independent}), this time including the results broken down by gender. For the \texttt{Gender-Dependent} approach, speech data from either male or female subjects was used to train and evaluate separate, gender-specific models. Column \texttt{All} shows, in each case, the aggregated results for all the subjects in the corpus. Interestingly, gender-dependent models do not lead to any performance improvement. The training data size reduction may explain the significant performance drop observed for the A-DNN models. On the contrary, our proposed gender-independent approach demonstrates strong robustness across genders. We hypothesize that, since the same speaker is present in both the ON and OFF states during training, the model is compelled to focus on subtle voice patterns or features --- relevant for the identification of PD medication state --- that are presumably independent of gender.

\noindent\textbf{Dysarthria Level Analysis.} Our analysis based on the speech-related MDS-UPDRS-III score did not reveal any clear pattern or consistent trend when examining model performance across dysarthria severity levels. This lack of correlation persisted even for subjects whose scores changed following medication. These observations align with prior studies suggesting that these scale ratings may not reliably capture speech-specific impairments \cite{vasquez2018towards}, thereby limiting their utility as benchmarks for speech-focused evaluations. The literature also highlights the inherent subjectivity and inter-rater variability often associated with these scale-based clinical assessments and their complexity \cite{richards1994interrater,ginanneschi1988evaluation}. In our particular case study, this is further compounded by ongoing debates regarding the effects of medication on speech, with some studies suggesting inconsistent or detrimental impacts on dysarthria severity \cite{quaglieri1977levotherapy}. Together, all these factors underscore the challenges and complexity of accurately assessing and interpreting dysarthria levels in clinical contexts.

\noindent\textbf{Comparison with Prior Work.} To the best of our knowledge, the work of Pompili et al. \cite{pompili2020medication} is the only approach for the automatic identification of PD medication state from speech. However, a direct comparison is not feasible, as their study approached this task from a speaker-dependent perspective and used a different task-grouping methodology. Nevertheless, in their best-performing setup, the authors reported an accuracy of 95.27\%, which can be considered reasonably close to the 88.2\% F\textsubscript{1}-score achieved here in the more challenging speaker-independent setting. Notably, their findings align with ours, showing improved performance in tasks involving continuous speech production. This shared observation underscores and further supports the critical importance of speech assessment task selection when addressing the identification of PD medication states, especially in speaker-independent scenarios.

\noindent\textbf{Limitations.} Beyond the limited data available and the associated risk of overfitting, one of the main limitations of this work is the no incorporation of demographic information, motor impairment severity scales, and personal treatment details ---factors considered in clinical assessments \cite{frota2021dyspros} and potentially beneficial for enhancing our model accuracy and applicability.

\section{Conclusions \& Future Work}

In this paper, we presented our work on the automatic identification of PD medication states from speech, showing its potential for more reliable remote monitoring and personalized treatment scheduling. Compared to prior work \cite{pompili2020medication}, our speaker-independent approach offers two key advantages: \textbf{i)} the ability to leverage larger, cross-speaker datasets, making it well-suited for data-scarcity scenarios, and \textbf{ii)} the avoidance of developing new models for each new individual patient.

Our experiments show robust performance across speakers and genders, suggesting the presence of gender-independent speech cues related to PD medication states. Pre-trained Wav2Vec2.0 features, extracted from continuous, prosody-rich speech tasks, emerge as the most effective approach for this task. Unlike the findings in \cite{pompili2020medication}, eGeMAPS features alone are insufficient in the speaker-independent setting. Additionally, attention-based architectures do not outperform traditional classifiers like SVMs. Our best-performing model achieves an F\textsubscript{1}-score of 88.2\%. The adoption of a task-specific training strategy further highlights the robustness of our approach in data-scarce environments, a common issue in pathological speech analysis. Moreover, our method's efficiency in requiring only one or a few speech assessments can reduce the burden on both patients and speech-language therapists in clinical settings.

Regarding future work, we plan to conduct a more systematic study on the relevance of prosody. While our experiments identified SSL-based representations as the most suitable approach for modelling continuous, prosody-rich speech tasks, the underlying mechanisms through which such information is encoded in these opaque deep-learning embeddings
remains unclear. In this context, our future work will pursue understanding models' decisions and investigating more interpretable SSL-based frameworks, as proposed in \cite{gimeno2025unveiling}. Incorporating demographic information, disease severity, and treatment details could further enhance the robustness of the approach, providing deeper insights into medication state analysis \cite{frota2021dyspros}.

\section{Acknowledgements}

The work of Gimeno-Gómez and Martínez-Hinarejos was partially supported by GVA through Grants CIACIF/2021/295 and CIBEFP/2023/167, by Grant PID2021-124361OB-C31 funded by MCIN/AEI/10.13039/501100011033 and ERDF/EU. The work of Solera-Ureña, Pompili and Abad was supported by Portuguese national funds through Fundação para a Ciência e a Tecnologia (FCT) under projects UI/DB/50021/2020 (doi: 10.54499/UIDB/50021/2020) and by the Portuguese Recovery and Resilience Plan and NextGenerationEU European Union funds under project C644865762-00000008 (Accelerat.AI). The FraLusoPark corpus collection was funded by the Agence Nationale de la Recherche (ANR) and FCT under project FCT-ANR/NEU-SCC/0005/2013.

\bibliographystyle{IEEEtran}
\bibliography{main}

\end{document}